\documentclass[11pt]{article}
\usepackage{amsmath,amsthm,amsfonts}
\usepackage{color}
\usepackage{graphicx}
\usepackage{vmargin}

\newtheorem{theorem}{Theorem}[section]
\newtheorem{lemma}[theorem]{Lemma}
\theoremstyle{definition}

\newtheorem{definition}[theorem]{Definition}

\newtheorem{proposition}[theorem]{Proposition}

\newtheorem{claim}[theorem]{Claim}

\newcounter{codelines}

\newcommand{\set}[1]{ \left \{ #1 \right \} }
\newcommand{\eproof}{\begin{flushright}$\Box$\end{flushright}}

\newcommand{\T}[3]{ T_{#1,#2}(#3)}
\newcommand{\dt}[3]{ \delta_{#1,#2}(#3)}
\newcommand{\R}[3]{ \mathcal{R}_{#1,#2}(#3)}

\def\A{{\mathcal{A}}}
\def\B{{\mathcal{B}}}

\def\P{{\mathrm{Pr}}}

\def\P{{\mathrm{Pr}}}
\def\E{{\mathrm{E}}}

\def\GSPPT{{\mathrm{GSPPT}}}
\def\GSPART{{\mathrm{GSPART}}}
\def\GWPPT{{\mathrm{GWPPT}}}
\def\GWPART{{\mathrm{GWPART}}}

\newcommand{\cclass}[1]{\mathbf{#1}}

\newcommand{\PAI}{\P_{(x,\sigma)}}
\newcommand{\PA}{\P_{\sigma}}

\title{Generic Case Complexity and One-Way Functions}

\date{}
\author{Alex D. Myasnikov}

\begin{document}

\maketitle

\begin{abstract}
The goal of this paper is to introduce ideas and methodology of the generic case complexity to cryptography community.
This relatively new approach allows one to analyze the behavior of an algorithm on ``most" inputs in a simple and
intuitive fashion which has some practical  advantages over classical methods based on averaging.

We present an alternative definition of one-way function using the concepts of generic case complexity  and show its equivalence to
the standard definition. In addition we demonstrate the convenience of the new approach by giving a short proof
that extending adversaries to a larger class of partial algorithms with errors does not change the strength of the security assumption.
\end{abstract}

\section{Introduction}

Generic case complexity has originated about a decade ago in combinatorial group theory \cite{KMSS1,BMR}. This area has long computational traditions
 with many fundamental problems being algorithmic in nature. It has been shown that most computational  problems
 in infinite group theory are recursively undecidable. However, it was also observed that  decision algorithms, sometimes very naive ones,
 exist for many inputs even if a problem is undecidable in general.

 Generic complexity was suggested as a way of analyzing the behavior of undecidable problems. The main question was
 to describe the complexity of a problem on a {\em generic} input or on a set which contains most of the inputs.
The idea was to separate sets of inputs where algorithms work from the ``bad" ones. It happened that quite often
inputs on which algorithms fail to provide an answer are small.

In computer science, around 1980s, the same kind of arguments preceded the development of the  average case complexity.
More recently, heuristic classes of algorithms were introduced \cite{BT}.

Advocates of generic complexity approach argue (see discussions in \cite{GMMU}) that it is simpler, intuitive and
more general then the average case complexity. The connection between the two areas has been studied and it is known
that there are problems which are hard on average, but generically easy.
It turns out however,  that if an algorithm is easy on average it is also easy generically.

The relation between generic complexity and heuristic complexity is less explored. It was shown \cite{GMMU} that the class of generic algorithms
and errorless heuristic algorithms are equivalent. It seems that generic complexity has some advantage as
the area has significantly progressed in recent years. For example the completeness theory for generic complexity has been developed.

Here we list some   results in generic complexity. As we mentioned above, the foundations were built
in  group theory. In particular it has been shown that the famous word and  conjugacy  problems in finitely presented groups can be decided in
linear time on a generic set of inputs, although these problems are undecidable in general \cite{KMSS1}.% Other developments include \cite{}.

In the scope of  the classical complexity results, the most important is the existence of polynomial
reductions for generic complexity. Using these reductions it has been shown that there exist generically NP-complete problems, for example bounded
versions of the halting and Post correspondence problems are generically NP-complete \cite{GMMU}.
Another interesting result shows that the halting problem for a model
of a Turing machine with  one-way infinite  tape is linearly decidable on a generic set of inputs \cite{HM}. It is not known whether the result holds for
an arbitrary Turing machine, but it was shown that the set on which the problems is decidable cannot be strongly generic \cite{R}.

In \cite{MR} authors describe a  particular procedure which allows one given an undecidable problem  to construct a problem undecidable on every generic set of inputs. This generic amplification shows  that generically hard (undecidable)  problems exist.

It was also suggested that generic complexity might be useful for cryptographic applications, particularly for  testing  security assumptions of
cryptographic primitives. Intuitively, we would like a cryptographic  primitive to be hard to break on most inputs which seems like a straightforward
application of the ideas of generic complexity. The main goal of this paper is to introduce ideas and methodology of generic complexity to
cryptography community. We present alternative  definitions of one-way functions based on  the concept  of generic complexity.

 These new definitions allow one to consider,  in a natural way,   one-way function candidates coming from undecidable problems.
We show  that any such ``generic" one-way function can be used to produce a classical one. Therefore, any new generic one-way function comes along with new classical one.
Furthermore, to our opinion these new  definitions are more intuitive  and are easier to work with. Indeed, the new security assumption is just a more precise  formalization of the original notion, due to Diffie and Hellman \cite{DH}, in a sense, it separates  the probability on the inputs from the probability on the oracle choices - which makes considerations  easier.  As an illustration,  we give a short proof
that extending adversaries to a larger class of partial algorithms with errors does not change the strength of the security assumption.

In the subsequent paper we are going to discuss some potential generic one-way functions that are related to undecidable problems in algebra.

\subsection{Generic complexity notations}

In this section we give a brief overview of the basic notions and definitions used in generic complexity.
 For more detailed introduction to the subject and latest results we refer to \cite{GMMU}.

Let $I$ be a set of inputs. In this paper we consider traditional binary representation of inputs and set $I = \set{0,1}^*$.
With each input we associate a size function $|\cdot| : I \rightarrow \mathbb{N}$ which
is the length of a string from $I$.

First we define  a {\em stratification} of inputs. In general a stratification
of the set $I$ is an ascending sequence of subsets whose union is equal to $I$. In the paper we will use the
spherical stratification on strings which we define next.

\begin{definition}[Spherical Stratification] Let $I= \set{0,1}^*$ be a set of inputs. Define a sphere of radius $n$ by
\[
I_n = \set{ x \mid x\in I, |x| = n}.
\]
Then the sequence $I_0, I_1, I_2, \ldots$ is a spherical stratification of $I$.
\end{definition}
Note that sets $I_i$ are finite and $\cup_{i=0}^{\infty} I_i = I$.

There are other commonly used stratifications available. For example one can stratify set $I$ using
balls $B_n$ of inputs of radius $n$, where $B_n$ is a set
of inputs with lengths at most $n$.

\begin{definition} Let $I = \set{0,1}^*$  and $I_n \subset I$ be a sphere of radius $n$. Let $\mu_n$ be a probability
distribution on the sphere $I_n$. The collection $\set{\mu_0, \mu_1, \mu_2, \ldots }$ of all distributions
is called  {\em an ensemble of spherical distributions over $I$ } and denoted by $\set{\mu_n}$.
\end{definition}

In the paper we will be mostly concerned with the ensemble of uniform spherical distributions $\set{u_n}$ over $I$. For
a set $R \subseteq I$ we define
\[
u_n(R) = \frac{|R \cap I_n|}{|I_n|},
\]
where $|X|$ is the cardinality of a set $X$.

Next we define an asymptotic density of a set in $I$.
\begin{definition}[Asymptotic Density] \label{def:assym_den} Let $\mu = \set{\mu_n}$ be an ensemble of spherical
 distributions over a set $I$. A set of inputs $R \subseteq I$ is said to have asymptotic density $\rho(R) = \alpha$
if
\[
\lim_{n\rightarrow \infty} \mu_n(R \cap I_n) = \alpha.
\]
 A set $R$ is called {\em generic} with respect to $\mu$ if its asymptotic density is 1 and it is called {\em negligible} if the
asymptotic density is 0.
\end{definition}

\begin{definition}
Let $R \subseteq I$ and the asymptotic density $\rho(R)$ exists. The function
\[
\delta_R(n) = \mu_n(R \cap I_n)
\]
is called the {\em density } function for $R$.
\end{definition}

A practical measure of the ``largeness" of a set often corresponds to a rate with which the limit in Definition \ref{def:assym_den} converges.
The convergence can be naturally described by obtaining upper bounds on the density function of a set.
One particular type of sets of interest are sets which have superpolynomial convergence rates.

\begin{definition} Let $R \subseteq I$ and $\delta_R(n)$ is the density function
of $R$. We say that $R$ has asymptotic density $\rho(R)$ with superpolynomial convergence if

\[
|\rho(R) - \delta_R(n)| < \frac{1}{p(n)}
\]
for every polynomial $p(n)$ and all sufficiently large $n$.
%\comment{Check what is the difference between this definition and the one in the survey: $ | \rho(R) - \delta_R(n)|$ is $o(n^{-k}) \forall k$.}
\end{definition}

\begin{definition}[Strongly Generic/Negligible]
A generic set with superpolynomial convergence is called {\em strongly} generic and its complement
is called a  strongly negligible set.
\end{definition}

\subsection{One-Way functions}

Existence of one-way functions is one of the most basic and important
assumptions in cryptography. In fact existence of one-way functions is a minimal assumption required for constructing other cryptographic primitives such as
pseudorandom number generators, encryption and signature schemes.

Diffie and Hellman \cite{DH} define one-way functions:
\begin{quote}

``a function $f$ is a one-way function if, for
any argument $x$ in the domain of $f$, it is easy to compute the corresponding
value $f(x)$, yet, for almost all $y$ in the range of
$f$, it is computationally infeasible to solve the equation $y = f(x)$
for any suitable argument $x$."
\end{quote}

There are
two key points in the definition above: ``for almost all" and ``computationally infeasible".
A lot of attention is still concentrated on the development
and understanding of these two notions and their consequences from the
practical point of view.

It is well accepted now that one-way functions cannot be defined using deterministic worst-case
complexity classes like $\cclass{P}$ and $\cclass{NP}$,
and   randomized computation
is the default model for cryptographic purposes.

A common argument for the necessary conditions for one-way functions
to exist proceeds as follows \cite{G}. Suppose we have a
cryptographic scheme. Legitimate parties should be able to decode
the secret efficiently, which means that there exist a
polynomial-time verifiable witness to the decoding and
 the problem of breaking a cryptographic scheme is in
$\cclass{NP}$. For a cryptographic scheme to be considered secure
there should be no practical algorithm to break the encryption.
Therefore, if a secure cryptographic scheme exists then $\cclass{NP}
\not \subseteq \cclass{BPP}$. Whether $\cclass{BPP}$ contains
$\cclass{NP}$  is an open problem. Note that $\cclass{NP} \not
\subseteq \cclass{BPP}$ implies that $\cclass{P} \neq \cclass{NP}$.

The $\cclass{NP} \not \subseteq \cclass{BPP}$ condition is a
necessary, but not sufficient condition for a secure cryptographic
scheme to exist. Observe that the probability distribution in the
definition of the class $\cclass{BPP}$ is taken over the internal
states of a probabilistic machine only. The condition which bounds
away the probability of an error must hold for all inputs.
  In this sense $\cclass{BPP}$ is analogues to
$\cclass{P}$ and is still reflects the behavior of a problem on the
worst case inputs but with respect to the randomized algorithms.

%\comment{add the arguments here } Our arguments for deficiencies of the worst case analysis from
%Section \ref{subsec:deficiency-worst} can be directly translated to
%the case of standard randomized classes. For instance,
The positive
answer to the problem $\cclass{NP} \not \in \cclass{BPP}$ may have
no practical implications for cryptography, unless there are
problems which belong in $\cclass{NP} \backslash \cclass{BPP}$ and
are hard on a significantly large fraction of inputs. Speaking in
terms of generic complexity, a problem may be considered hard if
there is no efficient algorithm which solves the problem on any but
strongly negligible set of inputs.

In cryptography the existence of many useful primitives like secure symmetric encryption, pseudorandom number generators and
 digital signature schemes is reduced to the
existence of the one-way functions which we define next. In general
there are two notions of one-way functions a strong and a weaker
one.

Let $\PAI$ denote the probability taken uniformly over all pairs $(x,\sigma) \in I_n \times \Sigma$, where
$I_n$ is the set of all inputs of length $n$ and $\Sigma = \{0,1\}^{t(n)}$ is the space of internal
coin flips of a probabilistic algorithm whose running time is bounded by some
polynomial $t(n)$. Similarly we define  $\PA$ as the uniform probability taken over $\Sigma$ only.

One of the most commonly accepted definitions of a one-way function (strong one-way function) is the following.
\begin{definition}[Strong One-Way function \cite{G}]\label{def:SOWFG2}
A function $f: \{0,1\}^* \rightarrow \{0,1\}^*$ is called strongly
one-way if the following two conditions hold:
\begin{enumerate}
\item Easy to compute: there exists a deterministic polynomial-time algorithm $\A'$ such that
on an input $x$ algorithm $\A'$ outputs $f(x)$;
\item Hard to invert: For every probabilistic polynomial-time algorithm $\A$, every positive
polynomial $p$, and all sufficiently large $n$:
\[
\PAI[A(f(U_n),1^n) \in f^{-1}(f(U_n))] < \frac{1}{p(n)},
\]
where $U_n$ is a random variable uniformly distributed over
$\{0,1\}^n$ and the probability is taken over
all input strings  from $\{0,1\}^n$ and internal states of $\A$.
\end{enumerate}
\end{definition}

Here and in the rest of the article {\em polynomial-time} algorithm means an algorithm 
that always halts after a polynomial (in the length of the input) number of steps.
Note that in addition to an input in the range of $f$ the algorithm $\A$ is given the auxiliary
input $1^n$ which has the same length as the desired output of $\A$. This is done to protect from 
the situations when the function $f$ drastically reduces the length of its input (for example $|f(x)| = \log_2(|x|)$).
Obviously no algorithm can invert such function $f$ in polynomial number of steps in terms of $|x|$.

\section{Generic definitions of one-way functions}
\label{sec:GEN_DEF}

\subsection{Definition restricted to PPT adversary}

In Definition \ref{def:SOWFG2} the performance of an algorithm $\A$ is averaged over all inputs
which results in complicated probability space. We would like to apply ideas of generic complexity
and consider the performance of an adversary on each input separately.

Note that a naive random sampling will guess an inverse of a function $f$ on
the input of length $n$ with probability $1/2^n$. An algorithm with negligible
probability of the correct answer cannot be amplified and, therefore, cannot be considered
practical. A reasonable inversion algorithm should have noticeable
probability of success. To be more precise the probability that an algorithm $\A$ inverts $f(x)$ 
\[
\P[A(f(x), 1^n) \in f^{-1}(f(x))] > \frac{1}{n^c}
\]
for any positive constant $c$.
To make a one-way function secure we must limit the number of
inputs on which adversary succeeds to a small set.
We formalize these arguments in the following definition of a generically strong one-way function.

\begin{definition}[Generically Strong One-Way function]
\label{def:SOWF2}
Let $u=\{u_n\}$ be an ensemble of uniform spherical distributions
over $\{0,1\}^*$.

A function $f: \{0,1\}^* \rightarrow \{0,1\}^*$ is called {\em generically strong
one-way} if the following two conditions hold:
\begin{enumerate}
\item Easy to compute: there exists a deterministic polynomial-time algorithm $\A'$ such that
on input $x$ algorithm $\A'$ outputs $f(x)$;
\item Hard to invert almost all inputs: For every probabilistic polynomial-time algorithm $\A$, all constants $c>0$, every positive
polynomial $p$ and  all sufficiently large
$n$: \[
u_n \left( \{ x \in I_n \mid \P[A(f(x), 1^n) \in f^{-1}(f(x))] > n^{-c} \} \right ) < \frac{1}{p(n)},
\]
where the probability is taken over internal states of the algorithm $A$.
\end{enumerate}
\end{definition}

Similarly we can define a generically weak one-way function.

\begin{definition}[Generically Weak One-Way function]
\label{def:WOWF2}
Let $u=\{u_n\}$ be an ensemble of uniform spherical distributions
over $\{0,1\}^*$.

A function $f: \{0,1\}^* \rightarrow \{0,1\}^*$ is called {\em generically weak
one-way} if the following two conditions hold:
\begin{enumerate}
\item Easy to compute: there exists a deterministic polynomial-time algorithm $\A'$ such that
on input $x$ algorithm $\A'$ outputs $f(x)$;
\item Hard to invert on a large enough set of inputs: For every probabilistic polynomial-time algorithm $\A$,
every constant $c>0$ there exists a polynomial $p(n)$ such that for all sufficiently large
$n$:
\[
u_n \left( \{ x \in I_n \mid \P[A(f(x),1^n) \in f^{-1}(f(x))] < n^{-c} \} \right ) \geq \frac{1}{p(n)},
\]
where the probability is taken over internal states of the algorithm $A$.
\end{enumerate}
\end{definition}

The following lemmas  show that definitions \ref{def:SOWF2} and \ref{def:SOWFG2} are equivalent.
We give equivalence results for strong one-way functions. Similar results hold for the weak notion as well (see Appendix for the detailed proof).
We use standard reduction argument which proceeds by showing  that if there exists an algorithms which violates the
conditions of the first definition then we can construct an algorithm which will violate conditions of
the second one.

\begin{lemma}\label{lem:PPT_SOWF_equiv1}
Let $f: \{0,1\}^* \rightarrow \{0,1\}^*$ and suppose there is a probabilistic polynomial time
algorithm $\A$ such that for some constants $c>0$ and $d>0$ and infinitely many $n$
\[
u_n \left (\{ x \in I_n \mid \PA[A(f(x),1^n) \in f^{-1}(f(x))] > n^{-c} \}\right) > \frac{1}{n^d}.
\]
Then there exists a probabilistic polynomial-time algorithm $\A'$  such that for infinitely many $n$
\[
  \PAI[A'(f(U_n),1^n) \in f^{-1}(f(U_n))] > \frac{1}{n^{d+1}}.
\]
\end{lemma}
{\em Proof.} First of all observe that since we can compute $f$, we can also check whether an algorithm indeed returns
an inverse of $f(x)$ or not. By definition, $f^{-1}(y) = \{ x \mid y = f(x)\}$ therefore if
$f(\A(f(x))) = f(x)$ then $\A(f(x))$ is an inverse of $f(x)$.

Now construct an algorithm $\A'$ as follows. Repeat algorithm $\A$ on a given input $x$ until a witness
for the inverse problem (i.e. the inverse itself) is obtained.
Let
\[
S_n = \left \{ x \in I_n \mid \PA[\A(f(x)) \in f^{-1}(f(x))] \geq n^{-c} \right \}.
\]
For the algorithm $\A'$ to be practical on the set $S_n$ we need to show that
for every $x \in S_n$ we can obtain an inverse with  high probability
using only polynomially many repetitions of $\A$, i.e.
\begin{equation}\label{eq:SOWFcond1}
\PA[\A_k'(f(x)) \in f^{-1}(f(x))] \geq 1 - \epsilon,
\end{equation}
where $k = p(n)$ and $\epsilon < \frac{1}{n^m}$ for any $m>0$.

%\mad{The following proof uses Chernoff bounds. It contains many details which should be disregarded perhaps}
Let $y_i$ be the output of the $i$th run of the algorithm $\A$ on an input $x \in S_n$ and
let $X_i$, $i=1,\ldots, k$ be random variables such that $X_i = 1$ if $y_i \in f^{-1}(f(x))$ and $X_i=0$ otherwise.
$X_i$ are mutually independent and $\E[X_i] = \P[X_i=1] \geq \frac{1}{n^c}$. We also
define $X^o_i$, $i=1,\ldots, k$ to be random variables such that $X^o_i = 0$ if $y_i \in f^{-1}(f(x))$
and $X_i^o=1$ if $i$th run of $\A$ fails.
$X^o_i$ are also mutually independent and $\E[X^o_i] = 1 - \P[X_i=1] \geq 1- \frac{1}{n^c}$.

Note for $\A'$ to produce an answer only one of $y_i$s needs to be a witness, therefore to show (\ref{eq:SOWFcond1})
we need to show that
\[
\P \left [ \sum_{i=1}^k X_i \geq 1 \right ] = \P \left [ \sum_{i=1}^k X^o_i \leq k-1 \right ] \geq 1 - \epsilon
\]
which is equivalent to showing
\[
\P \left [ \sum_{i=1}^k X^o_i > k-1 \right ] \leq \epsilon.
\]

Using Chernoff bound we have
\begin{eqnarray}
\P \left [\sum_{i=1}^k X_i^o - k\cdot \left (1-\frac{1}{n^c}\right) \geq \delta \cdot k\cdot\left (1-\frac{1}{n^c}\right) \right] &=& \\
\label{eq:SOWFcond2}= \P\left [\sum_{i=1}^k X_i^o \geq k\cdot\left(1-\frac{1}{n^c}\right)\cdot( \delta +1 )\right] &\leq& 2^{-\frac{\delta^2}{2}k}.
\end{eqnarray}
%To obtain $\delta$:
%\[
%k\cdot\left(1-\frac{1}{n^c}\right)\cdot( \delta +1 ) = k-1 \Rightarrow
%\]
%\begin{eqnarray*}
%\delta &=& \frac{n^c(k-1)}{k(n^c-1} -1 =.
%\end{eqnarray*}
Substituting $\delta = (k-n^c)/(k(n^c-1))$ into (\ref{eq:SOWFcond2}) we obtain
\[
\P\left[\sum_{i=1}^k X_i^o \geq k-1\right] \leq 2^{-\frac{1}{2}\cdot \left (\frac{k-n^c}{k(n^c-1)}\right )^2 k} = 2^{- \frac{(k-n^c)^2}{2k(n^c-1)^2}}.
\]
Let $k=n^{3c}$, then
\[
2^{- \frac{(k-n^c)^2}{2k(n^c-1)^2}} < 2^{- \frac{1}{2}(n+2)}
\]
and we have
\[
\P\left[\sum_{i=1}^k X_i^o \geq k-1\right] < 2^{- \frac{1}{2}(n+2)}.
\]
Therefore we obtained
\[
\PA[\A_k'(f(x)) \in f^{-1}(f(x))] \geq 1 - \epsilon,
\]
where $\epsilon = 2^{- \frac{1}{2}(n+2)}$. 
Note that a similar result can be obtained without using the Chernoff bound,
however, it allows us to obtain a tighter bound on the  number of repetitions of the algorithm $\A$.

 %\marginpar{\begin{am} Leave the Chernoff's proof in \end{am}}
% \mad{Alternative proof. Let $y_i$ be the output of the $i$th run of
% the algorithm $\A$ on an input $x \in S_n$ and let $X_i$,
% $i=1,\ldots, k$ be random variables such that $X_i = 1$ if $y_i \in
% f^{-1}(f(x))$ and $X_i=0$ otherwise. $X_i$ are mutually independent
% and $\P[X_i=1] \geq \frac{1}{n^c}$. Therefore the probability
% $\P[X_i=0] \leq 1 - \frac{1}{n^c}$. Note for $\A'$ to produce an
% answer  only one of $y_i$s needs to be a witness. The probability
% that we will not obtain an inverse after $k$ runs of the algorithm
% $\A$ is equal to $(1 - n^{-c})^k$. It is easy to see that there
% exists a polynomial $p(n)$ such that
% \[
% \PA[\A_{p(n)}'(f(x)) \not \in f^{-1}(f(x))]  \leq (1 - n^{-c})^{p(n)} < \frac{1}{2^n}.
% \]
% Hence we showed that there exists a polynomial algorithm $\A'$ such that
% \[
% \PA[\A'(f(x)) \in f^{-1}(f(x))]  \geq 1 - \frac{1}{2^n}.
% \]
% }

Taking the sum over all $x \in S_n$ we obtain
\[
\sum_{x\in S_n} \PA[\A'(f(x)) \in f^{-1}(f(x))] \geq \sum_{x\in S_n}(1 - \epsilon) = |S_n|(1 - \epsilon).
\]
Note that
\[
u_n(S_n) =\frac{|S_n|}{|I_n|} \geq \frac{1}{n^d}.
\]
Therefore
\[
|S_n| \geq \frac{|I_n|}{n^d} = \frac{2^n}{n^d}.
\]
It follows
\begin{eqnarray}\label{eq:SOWF_proof1}
\sum_{x\in S_n} \PA[\A'(f(x)) \in f^{-1}(f(x))] &\geq& |S_n|(1- \epsilon) \geq \frac{2^n}{n^d} \left ( 1 -  \epsilon \right ).
\end{eqnarray}

Next we show that $\PAI[\A'(f(U_n),1^n) \in f^{-1}(f(U_n))] \geq \frac{1}{n^d} - \epsilon$.

Define $A'(x,\sigma) = 1$ if the computation of  $\A'$
corresponding to oracle $\sigma$ inverts $f(x)$ and $A'(x,\sigma) = 0$ otherwise.

Now we have
\[
\PAI[\A'(f(U_n),1^n) \in f^{-1}(f(U_n))] = \sum_{\forall (x,\sigma)} %(x,\sigma) \in I_n \times \{0,1\}^{t(n)}}
A'(x,\sigma)p(x,\sigma),
\]
where $p(x,\sigma)$ is the joint probability mass function.

Note that $x$ and $\sigma$ are independent from each other, therefore
\begin{eqnarray*}
\sum_{\forall (x,\sigma)} A'(x,\sigma)p(x,\sigma) &=& \sum_{x \in I_n} \sum_{\sigma \in \{0,1\}^{t(n)}}A'(x,\sigma)p(x)p(\sigma)\\
&=& \frac{1}{2^n} \sum_{x\in I_n} \sum_{\sigma \in \{0,1\}^{t(n)}}A'(x,\sigma)p(\sigma)\\
&=& \frac{1}{2^n} \sum_{x \in I_n} \PA[\A'(f(x)) \in f^{-1}(f(x))].
\end{eqnarray*}

From (\ref{eq:SOWF_proof1}) and the equation above we have
\begin{eqnarray*}
\PAI[\A'(f(U_n),1^n) \in f^{-1}(f(U_n))] &=& \frac{1}{2^n} \sum_{x \in I_n} \PA[\A'(f(x)) \in f^{-1}(f(x))] \\
&\geq& \frac{1}{2^n} \sum_{x\in S_n} \PA[\A'(f(x)) \in f^{-1}(f(x))] \\
&=& \frac{1}{n^d}\left ( 1  - \epsilon \right ).
\end{eqnarray*}

Now let $d' = d+1$. It is easy to see that $1/n^d ( 1  - \epsilon) > 1/n^{d'}$ for $n \geq 2$.
%\begin{eqnarray*}
%\left( \frac{1}{n^d} -\epsilon \right ) - \frac{1}{n^{d'}} &=& \frac{n^{d+1} - n^d}{n^{d+1}n^d} - \epsilon \\
%&=&\frac{n-1}{n^{d+1}} - \epsilon \\
%&>& \frac{n}{n^{d+1}} - \epsilon \\
%&=& \frac{1}{n^d} - \epsilon \\
%&>& 0
%\end{eqnarray*}
%since $\epsilon$ is negligible.
Therefore we have
\[
\PAI[\A'(f(U_n),1^n) \in f^{-1}(f(U_n))] \geq \frac{1}{n^d}\left ( 1  -\epsilon \right )  > \frac{1}{n^{d+1}}.
\]

\eproof

The implication holds in the the opposite direction as well.
\begin{lemma}\label{lem:PPT_SOWF_equiv2}
Let $f: \{0,1\}^* \rightarrow \{0,1\}^*$ and suppose there is a probabilistic polynomial time
algorithm $\A$ such that for some polynomial $p(n)$ and infinitely many $n$
\[
\PAI[A(f(U_n),1^n) \in f^{-1}(f(U_n))] \geq  \frac{1}{p(n)}.
\]

Then there exists a probabilistic polynomial-time algorithm $\A'$  such that
for every $c>0$ and infinitely many  $n$
\[
u_n \left (\{ x \in I_n \mid \PA[A'(f(x)) \in f^{-1}(f(x))] > n^{-c} \}\right) \geq \frac{1}{2p(n)}.
\]
\end{lemma}
{\em Proof.} First we show that
\begin{equation}
u_n \left (\{ x \in I_n \mid \PA[A(f(x)) \in f^{-1}(f(x))] > 1/2p(n) \}\right) \geq \frac{1}{2p(n)}.
\end{equation}

The proof follows directly from the following averaging argument:
\begin{claim}
Let $a_1, \ldots, a_N \in [0,1]$ and $\rho \geq 0$ such that $\frac{1}{N}\sum_{i=1}^N a_i \geq \rho$
and let $k = \#\{ a_i \mid a_i > \rho/2 \}$. Then
\[
\frac{k}{N} \geq \frac{\rho}{2}.
\]
\end{claim}
%{\em Proof of the claim:} Note that $0 \leq \rho \leq 1$ and $1-\rho/2 > 0$. Then
%\[
%\frac{k}{n} + \left(1-\frac{k}{n}\right)\cdot \frac{\rho}{2} - \frac{\rho}{2} = %\frac{k}{n}\cdot \left ( 1 - \frac{\rho}{2} \right ) > 0.
%\]
%Therefore,
%\[
%\frac{k}{n} + \left(1-\frac{k}{n}\right)\cdot \frac{\rho}{2} > \frac{\rho}{2}
%\]
%and, since  $(1-\frac{k}{n}) > 0$
%\[
%\frac{k}{n} > \frac{k}{n} + \left(1-\frac{k}{n}\right)\cdot \frac{\rho}{2} > \frac{\rho}{2}.
%\]
%{\em End proof of claim}

Observe that
\[
\PAI[A(f(U_n),1^n) \in f^{-1}(f(U_n))] =
\frac{1}{2^n}\sum_{x \in I_n} \PA\left [A(f(x)) \in f^{-1}(f(x)) \right ]  \geq \frac{1}{p(n)}.
\]

If we set $a_i = \PA\left [A(f(x_i)) \in f^{-1}(f(x_i)) \right ]$, $x_i \in I_n$, $N = 2^n$,
$\rho = 1/p(n)$ and $k = \# \{ x \in I_n \mid \PA[A(f(x)) \in f^{-1}(f(x))] > 1/2p(n)\}$ then
it follows from the claim above that
\[
\frac{k}{2^n} \geq  \frac{1}{2p(n)}
\]
and
\[
u_n \left (\{ x \in I_n \mid \PA[A(f(x)) \in f^{-1}(f(x))] > 1/2p(n) \}\right) \geq \frac{1}{2p(n)}.
\]

 Now observe that for any $c>0$ there exists a probabilistic polynomial-time  algorithm $\A'$ such that
\begin{equation}\label{eq:OWF_equiv1}
 \#\{ x \in I_n \mid \PA[A'(f(x)) \in f^{-1}(f(x))] > n^{-c} \} \geq k.
\end{equation}
Indeed, in the case when  $n^{-c}  \geq 1/2p(n)$ the claim follows directly. In the second case when $n^{-c}  < 1/2p(n)$
we can use the probabilistic error reduction and construct an algorithm $\A'$ such that  (\ref{eq:OWF_equiv1}) holds.
Therefore there exists a polynomial-time  algorithm $\A'$ such that
\[
u_n \left (\{ x \in I_n \mid \PA[A'(f(x)) \in f^{-1}(f(x))] > n^{-c} \}\right) \geq \frac{1}{2p(n)}.
\]

\eproof

%\begin{definition}
%A function $\delta: \mathbb{N} \rightarrow \mathbb{N}$ is called
%noticeable if there exists a polynomial $p$ such that for large
%enough $n$
%\[
%\delta(n) > \frac{1}{p(n)}.
%\]
%\end{definition}

%\begin{definition}
%Let $\mu = \{ \mu_n\}$ be an ensemble of spherical distributions on
%$I$. A set $S \subset I$ is called a noticeable set with respect to
%spherical ensemble $\mu$ if its density function $\delta_S$ with
%respect to $\mu$ is noticeable.
%\end{definition}

The following result demonstrates the connection between the security
assumption and asymptotic properties of the input sets.
\begin{proposition}\label{pro:OW-notic}
A polynomial-time computable function $f: \{0,1\}^* \rightarrow
\{0,1\}^*$ is strongly one way if and only if every
probabilistic polynomial-time algorithm $\A$ fails to invert $f$ on
all but strongly negligible sets of inputs with respect to an ensemble of
uniform spherical distributions over $\{0,1\}^*$.
%, i.e.
%for any set $S \subset \{0,1\}^*$ such that its density function
%$\delta_S > 1/p(n)$ for some polynomial $p$
%\[
%\P[\A(f(x)) \in f^{-1}(f(x))] < 1/3.
%\]
\end{proposition}
{\em Proof.} Suppose $f$ is strongly one-way and
suppose there exists an algorithm $\A$ which inverts $f$ on a set $S$ which is not strongly negligible.
Then there exists a polynomial $p(n)$ such that
\[
u_n(\{ x\in I_n \mid \PA[\A(f(x)) \in f^{-1}(f(x))] > n^{-c} \}) = u_n(S \cap I_n)
= \delta_s(n) > \frac{1}{p(n)}.
\]
 Therefore $f$ is not strongly one-way by Definition \ref{def:SOWF2}.

Now, suppose $f$ is not one-way. Then there exists an algorithm $\A$ such
that
\[
u_n(\{ x \in I_n \mid \PA[A(f(x)) \in f^{-1}(f(x))] > n^{-c}\}) > \frac{1}{p(n)}
\]
for some polynomial $p$, which contradicts the proposition
assumption.

\eproof

\subsection{Generic definition with a more general adversary}
\label{sec:gen_part}
%Proposition \ref{pro:OW-notic} implies that the class of generic
%algorithms is not suitable enough for analysis of cryptographic
%problems. Indeed, consider an algorithm $\A$ which inverts function
%$f$ on a constant fraction of inputs $C_n \in I_n$ for any $n$, i.e.
%\[
%%u_n(\{ x \in I_n \mid \A(f(x)) \in f^{-1}(f(x))\}) = c, 0\leq c \leq 1.
%u_n( C_n ) = c,\  0\leq c \leq 1.
%\]
%
%Obviously  $C = \cup_{n \in \mathbb{N}}C_n$ is a non-negligible set
%and $\A$ has nothing to do with generic.

%Below we define a  more general class of partial algorithms with
%errors.
%
%\begin{definition}(Partial Algorithm with Errors)\label{def:PA_ERROR} Let $\mu=\{\mu_n\}$ be an ensemble
%of spherical distributions and $\D=(L,\mu)$ be a distributional
%problem. An algorithm $\A$ is a partial $\delta$-correct algorithm
%if the asymptotic density  of the set $S$ of inputs on which $\A$ is
%correct exists and
%\[
%\rho( S ) = \delta, 0 \leq \delta \leq 1,
%\]
%\end{definition}
%
%Partial algorithms which are 1-correct are the class of heuristic
%generic algorithms and if the halting set of a 1-correct partial
%algorithm is equal to the set $S$, then we obtain a standard generic
%algorithm.
%
%Definition \ref{def:PA_ERROR} introduces a very general class of algorithms which includes
%probabilistic algorithms and heuristic algorithms with errors.
%The class of partial algorithm may be considered as a practical model of an adversary.

 The most interesting question is whether the generic approach may give us new, more general security assumptions.
Note that the polynomial bound on the adversary is not necessary. The only condition
that a successful adversary needs to satisfy is to have an algorithm which terminates
in polynomial time and with correct answer on a non-negligible set of inputs.
Suppose we would like to make a security statement which holds against a much stronger adversary, i.e. a partial
probabilistic heuristic algorithm which may output incorrect answers. Although an adversary algorithm
may not terminate on some inputs, it would still be a threat if it
succeeds on a relatively large set of inputs.

\begin{definition}[Partial algorithm with errors] Let $I$ be the set of inputs.  We say that an algorithm $\A$ is a
partial algorithm with errors  if it is correct on a subset $X \subseteq I$ of inputs 
 and on the set $I - X$ it either does not stop or stops with an incorrect answer.
\end{definition}

To make a formal statement we need a notion of achievement ratio of an adversary which is similar to the notions given in \cite{GL,HILL}.

 \begin{definition}[Achievement ratio]  Let $f: \set{0,1}^* \rightarrow \set{0,1}^*$ be a function and let $\A$ be a partial probabilistic algorithm with errors.
 The achievement ratio of $\A$ on an instance $f(x)$ is defined as
\[
\R{\A}{f}{x} = \T{\A}{f}{x}/\dt{\A}{f}{x},
\]
where  $\T{\A}{f}{x}$ is the time required for $\A$ to terminate on the input $f(x)$ and
\[
\dt{\A}{f}{x} = \PA[\A(f(x),1^n) \in f^{-1}(f(x),1^n)].
\]
\end{definition}
Achievement ratio allows one to consider a larger class of algorithms whose running time may not be bounded by a polynomial. 
In order for an adversary to have a polynomial achievement ratio on a given  input $x$,  it has
to have both: the polynomial running time and a noticeable probability of inverting $f(x)$.

The following definition is an attempt to give an intuitive notion of a
generalized practical security assumption for a one-way function.

\begin{definition}
\label{def:SOWF_GENERAL}
Let $u=\{u_n\}$ be an ensemble of uniform spherical distributions
over $\{0,1\}^*$.

A function $f: \{0,1\}^* \rightarrow \{0,1\}^*$ is called strongly
one-way if the following two conditions hold:
\begin{enumerate}
\item Easy to compute: there exists a deterministic polynomial-time algorithm $\A'$ such that
on input $x$ algorithm $\A'$ outputs $f(x)$;
\item Hard to invert: For every partial probabilistic algorithm with errors $\A$, all constants $c > 0$, 
every positive polynomial $p$  and all sufficiently large $n$:
\[
u_n \left( \{ x \in I_n \mid \R{\A}{f}{x} \leq n^{c} \} \right ) < \frac{1}{p(n)}.
\]
\end{enumerate}
\end{definition}

%We believe this is the most general definition of a one-way function.

The   question is whether or not this definition gives us any advantage over the definitions given earlier.
The following argument says that if we allow only a polynomial number of steps for an adversary
on a success then, in fact, this definition is equivalent to the one which is limited to the PPT adversary.

The main idea is that since the  {\em success} of an adversary
on an input $x$ means that it has to terminate in polynomial number of steps,
then we do not really care if adversary is a partial algorithm or not. If we have a successful
partial algorithm then we can construct a PPT algorithm by allowing it to run for polynomial
number of steps and  this polynomial-time algorithm will be as successful as the partial one.

Let $\GSPPT$ and $\GSPART$ be the  classes of one way functions which satisfy conditions of Definition \ref{def:SOWF2} and Definition \ref{def:SOWF_GENERAL}
respectively.

\begin{proposition}\label{pro:SOW_equiv} A function $f \in \GSPPT$ if and only if $f \in \GSPART$.
\end{proposition}

{\em Proof.} First we show that $f \in \GSPART$ implies $f \in \GSPPT$. The 
proof is by contradiction. Let $f: \set{0,1}^* \rightarrow \set{0,1}^*$ and assume that $f \in GSPART$, but $f \not \in GSPPT$, then
there exists a PPT algorithm $A$, a constant $c> 0$, a polynomial $p(n)$ such that
for infinitely many $n$
\[
u_n(\{x \mid \delta_{A,f}(x) > n^{-c}\}) \geq \frac{1}{p(n)}.
\]

Note that  a PPT algorithm $A$ is also a partial probabilistic  algorithm such that
$T_{A,f}(x) \leq q(n)$, for some positive polynomial $q$ for all $x$. Therefore,

\begin{eqnarray*}
u_n(\{x \mid \delta_{A,f}(x) > n^{-c}\}) &\geq& \frac{1}{p(n)} \\
u_n(\{x \mid \delta_{A,f}(x)/T_{A,f}(x) > n^{-c} / T_{A,f}(x) \}) &\geq& \frac{1}{p(n)} \\
u_n(\{x \mid T_{A,f}(x) / \delta_{A,f}(x) <  n^c T_{A,f}(x) \}) &\geq& \frac{1}{p(n)} \\
u_n(\{x \mid \mathcal{R}_{A,f}(x)  <  n^c T_{A,f}(x) \}) &\geq& \frac{1}{p(n)} \\
u_n(\{x \mid \mathcal{R}_{A,f}(x)  \leq  n^d \}) &\geq& \frac{1}{p(n)}, \\
\end{eqnarray*}
where $d$ is chosen such that $n^d \geq q(n) \cdot n^c$.  This is a contradiction to
the condition $f \in GSPART$.

%if we set polynomial time
%bound $T(x) \leq p(|x|)$ for all $x$ and the lower bound on the success probability to $n^{-t}$, for some $t > 0$,
%then Definition \ref{def:SOWF_GENERAL} holds. This definition encapsulates all types of algorithms including partial algorithms with errors.

The proof in the opposite direction uses a similar argument.
% We contradict that f maybe a strong in PPT def, but not strong in GC def. i.e. the new def is not stronger
%be a length preserving function
Suppose that $f \in \GSPPT$ but $f \not \in \GSPART$. %not one-way in terms of Definition \ref{def:SOWF_GENERAL}.
In other words we suppose there exists a partial probabilistic  algorithm $\B$ such that for some polynomial $p(n)$ and infinitely many $n$
\[
u_n \left( \{ x \in I_n \mid \R{\B}{f}{x} \leq n^{c} \} \right ) \geq \frac{1}{p(n)}.
\]
%However for any probabilistic polynomial-time algorithm $\A$ and any polynomial $poly(n)$ there exists a constant $c>0$ such that
%for all large enough $n$
%\[
%u_n \left( \{ x \in I_n \mid \dt{\A}{f}{x} \geq n^{-d} \} \right ) < \frac{1}{poly(n)}.
%\]

 Define $\A$ to be an algorithm which on a given input $x \in I_n$ runs $\B$  for $n^c$ steps.

Let $S = \set{x \mid \R{\B}{f}{x} \leq n^c }$.  First observe that by the conjecture for all $x \in S$
\[
\dt{\B}{f}{x} \geq \frac{\T{\B}{f}{x}}{n^c} \geq \frac{1}{n^c}.
\]

Obviously, $\dt{\A}{f}{x} = \dt{\B}{f}{x}$ for all $x$ such that $\T{\B}{f}{x} \leq n^c$. Therefore, since $\dt{\B}{f}{x} \in [0,1]$
we have
\[
\dt{\A}{f}{x} = \dt{\B}{f}{x}
\]
 for all $x$ such that $\T{\B}{f}{x} \leq \dt{\B}{f}{x} \cdot n^c$, i.e. for all $x \in S$.

Hence we have $\dt{\A}{f}{x} \geq \frac{1}{n^c}$ for all $x \in S$ and
\[
u_n \left( \{ x \in I_n \mid \dt{\A}{f}{x} \geq n^{-c} \} \right ) \geq u_n(S) \geq \frac{1}{p(n)}.
\]

Therefore, a probabilistic polynomial time algorithm $\A$ inverts $f$ on a not strongly negligible set
which contradicts our assumption that $f$ is one-way with respect to Definition \ref{def:SOWF2}.

\eproof

Note that the proof is simple and quite compact. Using the equivalence
lemmas \ref{lem:PPT_SOWF_equiv1} and  \ref{lem:PPT_SOWF_equiv2} we can conclude that
the Definition \ref{def:SOWF_GENERAL} is equivalent to Definition \ref{def:SOWFG2} which is based
on the averaging argument. It seems that obtaining the same result would be a more difficult task
when working with the average type definitions directly.

Similarly one can define a weaker variation of a one-way function with a partial adversary.

\begin{definition} 
\label{def:WOWF_GENERAL}
Let $u=\{u_n\}$ be an ensemble of uniform spherical distributions
over $\{0,1\}^*$.

A function $f: \{0,1\}^* \rightarrow \{0,1\}^*$ is called weakly
one-way if the following two conditions hold:
\begin{enumerate}
\item Easy to compute: there exists a deterministic polynomial-time algorithm $\A'$ such that
on input $x$ algorithm $\A'$ outputs $f(x)$;
\item Hard to invert on non-negligible set: For every partial algorithm $\A$ and  every constant $c>0$,
 there exists a polynomial $p(x)$ such that for all sufficiently large $n$
\[
u_n \left( \{ x \in I_n \mid \R{\A}{f}{x} > n^{c} \} \right ) \geq \frac{1}{p(n)}.
\]
\end{enumerate}
\end{definition}

The equivalence result for weak one-way functions holds as well. Let $\GWPPT$ be the class of generically weak one-way
functions and $\GWPART$ be  the  class of one way functions satisfying
Definition \ref{def:WOWF_GENERAL}.

\begin{proposition}\label{pro:WOW_equiv} A function $f \in \GWPPT$ if and only if $f \in \GWPART$.
\end{proposition}

{\em Proof.} The proof is similar to the proof of Proposition \ref{pro:SOW_equiv}. Suppose
that $f \in GWPART$ but $f \not \in GWPPT$. Then there exists a PPT algorithm $\B$ and  constant $c > 0$
such that for all polynomials $p(n)$
\[
u_n(\{ x \mid \delta_{\B,f}(x) < n^{-c} \}) < \frac{1}{p(n)}
\]

The probabilistic polynomial time  algorithm $\B$ is a probabilistic partial algorithm such that its
time $T_{\B,f}(x) \leq q(n)$ for some positive polynomial $q$ and all $x$.

Therefore, there exists a probabilistic partial algorithm $\B$ such that for all positive polynomials
$p$:
\begin{eqnarray*}
\frac{1}{p(n)} &>& u_n(\{ x \mid \delta_{\B,f}(x) < n^{-c} \})  \\
&=&  u_n(\{ x \mid T_{\B,f}(x)\delta_{\B,f}(x) < T_{\B,f}(x) n^{-c} \})  \\
&=& u_n(\{ x \mid T_{\B,f}(x)/ \delta_{\B,f}(x) \geq   T_{\B,f}(x) n^{c} \})  \\
&=& u_n(\{ x \mid R_{\B,f}(x) \geq  T_{\B,f}(x) n^{c} \})  \\
&\geq& u_n(\{ x \mid R_{\B,f}(x) \geq  n^d, \forall d>0 \})  \\
\end{eqnarray*}

Which contradicts the assumption that $f \in GWPART$.

Now note that
if $f$ is not weakly one-way in terms of Definition \ref{def:WOWF_GENERAL} then there exists a partial algorithm $\B$
such that for some constant $c > 0$ and every polynomial $poly(n)$
\[
u_n \left( \{ x \in I_n \mid \R{\B}{f}{x} \leq  n^{c} \} \right ) \geq 1 - \frac{1}{poly(n)}.
\]
Define a probabilistic polynomial-time algorithm $\A$ which runs $\B$ for $n^c$ steps.
Using the equalities from Proposition \ref{pro:SOW_equiv} we obtain
\[
u_n \left( \{ x \in I_n \mid \dt{\A}{f}{x} \geq  n^{-c} \} \right ) \geq u_n \left( \{ x \in I_n \mid \R{\B}{f}{x} \leq  n^{c} \} \right ) \geq 1 - \frac{1}{poly(n)}.
\]
Therefore,
\[
u_n \left( \{ x \in I_n \mid \dt{\A}{f}{x} <  n^{-c} \} \right )  < \frac{1}{poly(n)}
\]
for any polynomial $poly(n)$. Therefore,  $f$ is not weakly one way with respect to a PPT algorithm $\A$.

\eproof

One of the important results about one-way functions is the  so-called amplification theorem which states
that having a weak one-way function we can always construct a strong one.
Equivalences shown above allow us to make a similar statement for generic one-way function.

\begin{theorem}[Amplification] Generically weak one-way functions exist if and only if generically strong one-way
functions exist.
\end{theorem}

{\em Proof.} The proof is a corollary of the equivalence  Lemmas \ref{lem:PPT_SOWF_equiv1}, \ref{lem:PPT_SOWF_equiv2}, \ref{pro:SOW_equiv}, \ref{pro:WOW_equiv}
and the classical amplification theorem.

\eproof

\section{Conclusion}
The definition based on generic case complexity methodology has significant advantage
in the fact that the probabilities over inputs and internal states of the algorithm are taken separately.
 The definition is very intuitive and easy to understand. In fact it may be seen as a direct formalization of the definition
by Diffie and Hellman which we quote in the introduction.

Operating with simpler probability spaces and considering inputs separately may have some practical implications.
The work in this direction started very recently and the potential of generic approach has been little realized.
It would be interesting to see if generic complexity can be used to simplify definitions of cryptographic primitives and reducibility arguments.
Applications of generic case complexity analysis of the security of particular one-way function
candidates is also could be of great interest.
%
%
%
%it is important to mention that, despite a huge research interest and amount of work, it is still not known whether one-way functions
%exist.  There are well known candidates such as integer factorization or discrete log which believed to be hard but proofs yet to be
% discovered.  It seems that the new definition of one-way functions may broaden the pool of problems that we can use.
%
%The following questions may be of interest:
%\begin{question} Do  one-way functions in terms of definition \ref{def:SOWF_GENERAL} exist.
%\end{question}
%
%
%\begin{question} Does existence of one-way functions in terms of definition \ref{def:SOWF_GENERAL} implies the existence
%of one-way functions in terms of definition  \ref{def:SOWFG2}.
%\end{question}
%

\appendix

\section{Proof of equivalence for the definitions of the Weak One-Way functions}

The following is the classical definition of a weak one-way function.
\begin{definition}[Weak One-Way function]\label{def:WOWF}
A function $f: \{0,1\}^* \rightarrow \{0,1\}^*$ is called weakly
one-way if the following two conditions hold:
\begin{enumerate}
\item Easy to compute: there exists a deterministic polynomial-time algorithm $\A'$ such that
on an input $x$ algorithm $\A'$ outputs $f(x)$;
\item Slightly hard to invert: There exists a polynomial $p$ such that for every PPT $\A$ and all sufficiently large $n$:
\[
\PAI[\A(f(U_n),1^n) \not \in f^{-1}(f(U_n))] \geq  \frac{1}{p(n)},
\]
where $U_n$ is a random variable uniformly distributed over
$\{0,1\}^n$ and the probability is taken over
all input strings  from $\{0,1\}^n$ and internal states of $\A$.
\end{enumerate}
\end{definition}

\begin{proposition} Definitions  \ref{def:WOWF} and \ref{def:WOWF2} are equivalent.
\end{proposition}
The following two lemmas give the proof.
 Denote
\[
\delta_{\A,f}(x) = \P[A(f(x),1^n) \in f^{-1}(f(x))]
\]
and
\[
\bar{\delta}_{\A,f}(x) = \P[A(f(x),1^n) \not \in f^{-1}(f(x))].
\]
Obviously
\[
\delta_{\A,f}(x) = 1 - \bar{\delta}_{\A,f}(x).
\]
\begin{lemma}[Generic implies Classic] Suppose there exists a PPT algorithm $\A$ such that for some (equivalently all) $c > 0$,  all polynomials $p$
and infinitely many $n$
\[
u_n \left( \{ x \in I_n \mid \delta_{\A,f}(x) < n^{-c} \} \right ) < \frac{1}{p(n)}
\]
then there exists a PPT algorithm $\A'$ such that for all polynomials $q(n)$ and infinitely many $n$
\[
\PAI[\A(f(U_n),1^n) \not \in f^{-1}(f(U_n))]  < \frac{1}{q(n)}.
\]
\end{lemma}
{\em Proof.}

Observe that
\[
u_n(\{ x \mid \delta_{\A,f}(x) \geq n^{-c} \}) \geq 1 - \frac{1}{p(n)}.
\]

Let
\[
S_n = \{ x \mid \delta_{\A,f}(x) \geq n^{-c} \}
\]
Then
\[
u_n(S_n) = \frac{|S_n|}{2^n} \geq 1 - \frac{1}{p(n)}
\]
However,
\[
\sum_{x \in S_n} \delta_{\A,f}(x) \geq \sum_{x \in S_n} n^{-c} = |S_n| n^{-c}
\]
and we obtain
\[
\frac{1}{2^n} \sum_{x \in S_n}\delta_{\A,f}(x) \geq \frac{|S_n|}{2^n}n^{-c} \geq n^{-c}\left ( 1-\frac{1}{p(n)} \right ) = \frac{p(n) - 1}{n^c p(n)} > \frac{1}{n^c p(n)}
\]

From the proof of   the equivalence for the case of strong one way functions  we know that
\[
\PAI[\A(f(U_n),1^n) \in f^{-1}(f(U_n))] \geq \frac{1}{2^n} \sum_{x \in S_n} \delta_{\A,f}(x).
\]
Therefore
\[
\PAI[\A(f(U_n),1^n) \in f^{-1}(f(U_n))] \geq \frac{1}{n^c p(n)}
\]

Again, from the proof of the strong version we know that
by repeating the algorithm $\A$ polynomially many times we can obtain an algorithm $\A'$ such that
\[
\PAI[\A'(f(U_n),1^n) \in f^{-1}(f(U_n))] \geq 1 - \epsilon
\]
where $\epsilon < 1/q(n)$ for any positive polynomial $q(n)$. Then
\[
\PAI[\A'(f(U_n),1^n) \not \in f^{-1}(f(U_n))] < 1 - ( 1 -\epsilon) = \epsilon < \frac{1}{q(n)}
\]
for all polynomials $q(n)$.

\eproof

\begin{lemma}[Classic implies Generic] Suppose there exists a PPT algorithm $\A$ such that  for all polynomials $p$
and infinitely many $n$
\[
\PAI[\A(f(U_n),1^n) \not \in f^{-1}(f(U_n))]  < \frac{1}{p(n)}.
\]

then there exists a PPT algorithm $\A'$ such that for some (equivalently all) $c > 0$, all polynomials $p(n)$ and infinitely many $n$
\[
u_n \left( \{ x \in I_n \mid \delta_{\A',f}(x) < n^{-c} \} \right ) < \frac{1}{p(n)}
\]

\end{lemma}
{\em Proof.}
Let
\[
S_n = \{x \mid \bar{\delta}_{\A,f}(x) \geq n^{-d}\}
\]

Observe that
\[
n^d \cdot \PAI[\A(f(U_n),1^n) \not \in f^{-1}(f(U_n))] < \frac{1}{p(n)}
\]
for all positive polynomials $p(n)$.

{\em Proof.} Suppose that it is not. Then there exists a polynomial $p'(n)$ such
that
\[
n^d \cdot \PAI[\A(f(U_n),1^n) \not \in f^{-1}(f(U_n))] \geq \frac{1}{p'(n)}
\]
and
\[
\PAI[\A(f(U_n),1^n) \not \in f^{-1}(f(U_n))] \geq \frac{1}{p'(n)n^d}
\]
which contradicts the condition of the lemma.

Now using the same argument as in the previous proofs we can show that
\[
\PAI[\A(f(U_n),1^n) \not \in f^{-1}(f(U_n))] = \frac{1}{2^n} \sum_{x \in I_n} \bar{\delta}_{\A,f}(x) \geq \frac{1}{2^n} \sum_{x \in S_n} \bar{\delta}_{\A,f}(x)
\]
Therefore, for every $p(n)$
\begin{eqnarray*}
\frac{1}{p(n)} &>& n^d \cdot \PAI[\A(f(U_n),1^n) \not \in f^{-1}(f(U_n))] \\
 &\geq& \frac{n^d}{2^n} \sum_{x \in S_n} \bar{\delta}_{\A,f}(x) \\
 &\geq& \frac{n^d}{2^n} \sum_{x \in S_n} n^{-d}  \\
 &=& n^d \cdot \frac{|S_n|}{2^n} \cdot n^{-d} \\
 &=& u_n(S_n).
\end{eqnarray*}

Note that
\[
S_n = \{ x \mid 1 - \bar{\delta}_{\A,f}(x) < 1 -  n^{-d}\}\} =  \{ x \mid \delta_{\A,f}(x) < 1 -  n^{-d}\}\}.
\]

Using amplification we can construct a PPT algorithm $\A'$ which repeats $\A$ polynomially many times and such that
\[
S_n = \left \{ x \mid \delta_{\A',f}(x)  < \frac{1}{2^n} \right \}
\]

Therefore, there exists a PPT algorithm $\A'$ such that for every polynomial $p(n)$
\[
u_n \left ( \left \{ x \mid \delta_{\A',f}(x)  < \frac{1}{2^n} \right \} \right ) < \frac{1}{p(n)}.
\]
\eproof


\begin{thebibliography}{\hspace{0.5in}}

\bibitem{BT}
A.Bogdanov and L.Trevisan, {\it Average-Case Complexity}, Now Publishers Inc, 2006.

\bibitem{BMR}
A.V.Borovik, A.G.Miasnikov and V.N.Remeslennikov. {\it Multiplicative measures on free groups}, Internat. J. Algebra Comp., 13 no. 6 (2003), 705-731.
\bibitem{G}
O.Goldreich, {\it Foundations of cryptography}, Cambridge University Press, 2001.
\bibitem{DH}
W.Diffie and M.Hellman, {\em New Directions in Cryptography},
IEEE Transactions on Information Theory, V. IT-22, no. 6 (1976), 644--654.
\bibitem{GMMU}
R.Gilman, A.G.Miasnikov, A.D. Myasnikov and A. Ushakov. {\em Generic complexity of algorithmic problems},
Preprint, 2007.
\bibitem{GL}
O.Goldreich and L.Levin, {\em A hard-core predicate for all one-way functions,}
Proceedings of the twenty-first annual ACM symposium on Theory of computing, (1989), 25 -- 32.
\bibitem{GM}
S. Goldwasser and S. Micali. {\em Probabilistic Encryption}, JCSS,
28, 2 (1984), 270--299.
\bibitem{HILL}
J. Hastad, R. Impagliazzo, L. Levin and M. Luby, {\em Construction of Pseudorandom Generator from any One-Way Function,}
Manuscript, 1993.
\bibitem{HM}
J.D.Hamkins, A.G.Miasnikov. {\em The halting problem is decidable
on a set of asymptotic probability one.} Notre Dame J. Formal Logic Volume 47, Number 4 (2006), 515-524.

\bibitem{KMSS1}
I.Kapovich, A.G.Miasnikov, P.Schupp, V.Shpilrain, {\em Generic-case complexity, decision problems in group theory and random walks,} J. Algebra 264 (2003), 665-694.
%\bibitem{KMSS2}
%I. Kapovich, A. G. Myasnikov, P. Schupp, V.Shpilrain, {\em Average-case complexity and decision problems in group theory,}
%Advances in Math. 190 (2005), 343-359.
\bibitem{MR}
A. Miasnikov, A. Rybalov, {\em On Generically Undecidable Problems},
Preprint, 2007.
\bibitem{Papa}
C.~Papadimitriou, \emph{Computation Complexity}, (1994),
Addison-Wesley.
\bibitem{R}
A.Rybalov. {\em On the Strongly Generic Undecidability of the Halting
Problem}.  Theor. Comput. Sci. 377(1-3) (2007), 268-270
\end{thebibliography}
\end{document}